\documentclass[english]{article}
\usepackage{setspace}
\doublespace
\usepackage[T1]{fontenc}
\usepackage[latin9]{inputenc}
\usepackage{geometry}
\usepackage[active]{srcltx}
\usepackage{units}
\usepackage{amsmath}
\usepackage{amssymb}
\usepackage{graphicx}
\usepackage[authoryear]{natbib}

\makeatletter
\usepackage{url}
\usepackage{natbib}
\usepackage[font=small,format=plain,labelfont=bf, margin=0.5cm]{caption}
\usepackage{hyperref}
\hypersetup{colorlinks=true,citecolor=blue}
\date{}

\makeatother

\usepackage{babel}
\begin{document}

\title{\bf {Non-Smooth Variational Data Assimilation with Sparse Priors}}

\author{A.M. Ebtehaj\textsuperscript{1,2}, E. Foufoula-Georgiou\textsuperscript{1},
S.Q. Zhang\textsuperscript{3} and A.Y. Hou\textsuperscript{3}\\
{\small \textsuperscript{1}Department of Civil Engineering, Saint
Anthony Falls Laboratory, University of Minnesota}\\
{\small \textsuperscript{2}School of Mathematics, University of Minnesota}\\
{\small \textsuperscript{3}NASA Goddard Space Flight Center, Greenbelt,
Maryland}\\
 \hrulefill}

\maketitle
This paper proposes an extension to the classical 3D variational data
assimilation approach by explicitly incorporating as a prior information,
the transform-domain sparsity observed in a large class
of geophysical signals. In particular, the proposed framework extends
the maximum likelihood estimation of the analysis state to the maximum
a posteriori estimator, from a Bayesian perspective. The promise of the methodology
is demonstrated via application to a 1D synthetic example.

\section{Introduction }

Data Assimilation has played a central role in improving the forecasting
ability of hydro-meteorologic, climatic and oceanic modeling systems.
The basic idea is to consistently fuse the observations of the prominent
state variables (e.g., wind, temperature, pressure) or physical states
(e.g., cloud moisture, precipitation), iteratively in time,
into the knowledge of a Numerical Prediction Model (NPM) to reduce the
estimation uncertainty of the state variables of interest.
Data assimilation methods typically use the observations
to update the current a priori model estimates of the states (\emph{background})
and produce an a posteriori state (\emph{analysis}) to be used for\emph{
}prediction of the next time step (\emph{forecast}).

Data assimilation primarily stems from the least-squares estimation in
the statistical sense. Basically, the available methods can be divided into two
major categories. The first, is the group of recursive (least-squares)
\emph{filtering} methods which essentially exploit the temporal evolution
of some statistical characteristics of the system (e.g., covariance)
to efficiently track the optimal states sequentially in time (e.g.,
Kalman filter driven data assimilation methods). The second category,
the so-called \emph{variational} methods, relies on a batch
mode direct optimization at each instant of time when the observations
become available (e.g., 3D or 4D variational approaches). We remark
here that, these two apparently distinct approaches often share similar
mathematical concepts and are quite equivalent in many cases;
however, with different implementation strategies in practice. In
this paper, we restrict our attention to the second group of assimilation
methods and in particular the more primitive 3D variational (3D-VAR)
formulation. For a thorough review of the historical evolution of
the data assimilation techniques the reader is referred to \citet{[TalC87]}, \citet{[GhiM91]}, \citet{[Dal93]}, \citet{[BouC02]}, \citet{[Kal03]}, \citet{[ZhoEM05]}, \citet{[Eve09]},
and references therein.

The classical variational data assimilation typically involves solving
a \emph{smooth} optimization problem in which the solution has a minimum
weighted Euclidean distance to both observation and background estimates
where the weights are dictated by the pair of model and observation
error covariance matrices. From a statistical estimation point of
view this procedure is equivalent to the Maximum Likelihood (ML) estimation
of the unknown state in a Gaussian noise (error) environment. In this
classical formulation no \emph{a priori} assumption is explicitly
taken into account about the underlying structure of the analysis
state.

Natural signals can typically be projected onto transform domains
(e.g., Fourier, Discrete Cosine, Wavelet) in which a large fraction
of the representation coefficients is very close to zero and only
a few of them are significant, a signature typically referred to as
 ``sparsity''. For instance, the wavelet transform of piece-wise
smooth natural signals with occasional rapid variations often translates
to non-Gaussian heavy tail distribution of the wavelet coefficients
with a concentrated probability mass around zero.

Here, we propose a new formalism for variational data assimilation
which explicitly incorporates the underlying sparsity in the analysis
state as an a priori knowledge. In a very simple example we demonstrate
how this a priori knowledge can stabilize and make the computation of the
analysis state more accurate compared to a classical solution.

Section 2 is devoted to explaining the notation. In Section 3, we
briefly review the preliminary concept of the 3D-VAR data assimilation
scheme. In this setting, an elementary 1D example in the Gaussian
domain is presented to elaborate on the underlying assumptions and
performance of the methodology in an ideal case. In Section 4, we
provide evidence on the sparsity of some important geophysical signals.
Exploiting the observed sparsity as an a priori knowledge, in Section
5, we cast the variational data assimilation in a Bayesian framework
both in the spatial and wavelet domains. The promise of the methodology
is demonstrated through an elementary constructed 1D example in that
section. Section 6, contains a brief discussion and points out to
future research.

\section{Notation}

We refer to a 2D signal ${\rm X}$ as a vector $\mathbf{x}$, by stacking
all the pixels in a fixed order. All vectors are column vectors and
$\left(\cdot\right)^{T}$ indicates the transpose. For any vector
$\mathbf{x}\in\mathbb{R}^{n}$, $x_{i}$ refers to its $i^{th}$ element,
where $i\in\left\{ 1,\ldots,n\right\} $. The same notation applies
to a matrix operator $\mathbf{H}$ and its entries $h_{i,i}$. The
standard $l_{p}$-norm of $\mathbf{x}$ is denoted by $\left\Vert \mathbf{x}\right\Vert _{p}:=\left(\sum_{i=1}^{n}\left|x_{i}\right|^{p}\right)^{1/p}$,
for $p\geq1$, while the infinity norm is $\left\Vert \mathbf{x}\right\Vert _{\infty}=\max_{i}\left|x_{i}\right|$.
For $p<1$, $\left\Vert \mathbf{x}\right\Vert _{p}$ is no longer
a norm and hence not convex; nevertheless, we will use the term norm
in this case as well, keeping in mind this reservation. By weighted
inner product, we denote $\left\langle \mathbf{x},\mathbf{y}\right\rangle _{\mathbf{P}}:=\mathbf{x}^{T}\mathbf{P}\mathbf{y}$
and hence the corresponding weighted Euclidean (quadratic) norm, is
$\left\Vert \mathbf{x}\right\Vert _{\mathbf{P}}=(\mathbf{x}^{T}\mathbf{P}\mathbf{x})^{\nicefrac{1}{2}}$,
where $\mathbf{P}\succ0$ is a symmetric positive definite matrix.
In a linear transformation $\mathbf{\Phi}=\left[\phi_{1},\phi_{2},\ldots,\phi_{m}\right]\in\mathbb{R}^{n\times m}$
with $m\geq n$ (e.g., wavelet transform), the columns $\phi_{i}\in\mathbb{R}^{n}$
denote the ``atoms'' whereby any certain class of signals $\mathbf{x}\in\mathbb{R}^{n}$
can be well approximated by a linear combination of $\phi_{i}$'s,
i.e., $\mathbf{x}\cong\sum_{i}^{m}\phi_{i}c_{i}=\mathbf{\Phi}\mathbf{c}$.
The vector $\mathbf{x}$ is said to exhibit a sparse representation
in $\mathbf{\Phi}$, if the number of (significantly) non-zero elements
of the representation coefficients $\mathbf{c}$, is much smaller
than the signal dimension.

\section{Variational Data Assimilation}

The theory of (recursive) least-squares estimation has been central
to the development of the classical data assimilation methodologies.
Let the true state of interest at time $t$ be denoted by $\mathbf{x}(t)\in\mathbb{R}^{m}$,
a noisy observation of the state by $\mathbf{y}(t)\in\mathbb{R}^{n}$,
and the background estimate of the state produced by a dynamical model
of the underlying physics by $\mathbf{x}_{b}(t)\in\mathbb{R}^{m}$.
We assume that the model can reproduce an unbiased but noisy estimate
of the true state. In other words, it is assumed that the model can
resolve the underlying physics in such a way that under a consistent
perturbation of the input parameters and states, the ensemble average
of the output tends to the true state as the number of ensemble members
goes to the infinity, whilst the random deviation of each ensemble
member can be well approximated by a Gaussian density. In a more formal
setting we have two equations that relate the true state to the background
state and observation as follows:
\begin{align}
\mathbf{x}_{b} & =\mathbf{x}+\mathbf{w}\nonumber \\
\mathbf{y}\, & =\mathcal{H}(\mathbf{x})+\mathbf{v},\label{eq:1}
\end{align}
where $\mathbf{w}\sim\mathcal{N}\left(0,\,\mathbf{B}\right)$, $\mathbf{v}\sim\mathcal{N}\left(0,\mathbf{\, R}\right)$
are uncorrelated Gaussian and the time index is dropped for brevity.

Obviously the goal is now to obtain the so-called analysis state $\mathbf{x}_{a}\in\mathbb{R}^{m}$
as the best estimate of the true state, given the above pair of observation
and the background state. From the 3D variational point of view, it amounts to obtaining the analysis
state $\mathbf{x}_{a}$ which minimizes the sum of two quadratic cost
functions, each of which quantifies the weighted Euclidean distance
of the analysis to the background state $\mathbf{x}_{b}$ and observation
$\mathbf{y}$:
\begin{equation}
\mathbf{x}_{a}=\underset{\mathbf{x}}{{\rm argmin}}\left\{ \frac{1}{2}\left\Vert \mathbf{x}_{b}-\mathbf{x}\right\Vert _{\mathbf{B}^{-1}}^{2}+\frac{1}{2}\left\Vert \mathbf{y}-\mathcal{H}(\mathbf{x})\right\Vert _{\mathbf{R}^{-1}}^{2}\right\} ,\label{eq:2}
\end{equation}
where the weights are inverse of the error covariance matrices. For
now, we assume that the measurement operator $\mathcal{H}(\cdot)$
can be replaced with a linear time invariant $\mathbf{H}\in\mathbb{R}^{n\times m}$
operator. In the context of our study, the incremental formulation
for nonlinear measurement operator, see \citep[e.g., ][]{[CouTH94]}, which typically
arises in direct assimilation of satellite radiance observations,
will be briefly discussed in Section 5 .

From a statistical estimation point of view, the variational form
in equation\,(\ref{eq:2}) can be obtained through an ML estimator,
$\mathbf{x}_{ML}=\arg\max_{\mathbf{x}}\, p\left(\mathbf{y,}\mathbf{x}_{b}|\mathbf{x}\right)$,
where $p\left(\mathbf{y,}\mathbf{x}_{b}|\mathbf{x}\right)$ denotes
the joint conditional density (Gaussian) of the observation and background
state with respect to $\mathbf{x}$. In this view, the cost function
in equation (\ref{eq:2}) is equivalent to the negative of the log-likelihood
function, assuming the background and observation vectors are independent,
see \citep[e.g., ][]{[BouC02]}. Simple algebra and ignoring the constant
terms in $\mathbf{x}$ yields a smooth quadratic cost function
\begin{align}
\mathcal{J}(\mathbf{x})= & \frac{1}{2}\mathbf{x}^{T}\left(\mathbf{B}^{-1}+\mathbf{H}^{T}\mathbf{R}^{-1}\mathbf{H}\right)\mathbf{x}-\nonumber \\
 & \left(\mathbf{B}^{-1}\mathbf{x}_{b}+\mathbf{H}^{T}\mathbf{R}^{-1}\mathbf{y}\right)^{T}\mathbf{x},\label{eq:3}
\end{align}
where the analysis state is its potential unique minimizer, $\mathbf{x}_{a}=\arg\min_{\mathbf{x}}\,\mathcal{J}(\mathbf{x)}$.
Note that the cost function of equation\,(\ref{eq:3}) is strictly
convex with unique global minimum provided that the Hessian $\nabla^{2}\mathcal{J}(\mathbf{x})=\mathbf{B}^{-1}+\mathbf{H}^{T}\mathbf{R}^{-1}\mathbf{H}$
is positive definite which requires that the measurement operator
$\mathbf{H}$ be a full rank matrix, see \citet[p.160]{[OlvS06]}.
This unique minimum can be obtained by setting the first order derivative
to zero $\nabla_{\mathbf{x}}\mathcal{J}=0$
\begin{equation}
\mathbf{x}_{a}=\left(\mathbf{B}^{-1}+\mathbf{H}^{T}\mathbf{R}^{-1}\mathbf{H}\right)^{-1}\left(\mathbf{B}^{-1}\mathbf{x}_{b}+\mathbf{H}^{T}\mathbf{R}^{-1}\mathbf{y}\right).\label{eq:4}
\end{equation}
Through the Fisher information it can be shown that the obtained analysis
in equation\,(\ref{eq:4}) can be an \emph{efficient} (unbiased with minimum variance)
estimator and its error covariance meets the \emph{Cramer-Rao}
lower bound, which is the inverse of the Hessian of $\mathcal{J}(\mathbf{x})$,
see \citep[e.g.,][p. 140]{[BouC02],[Lev08]}.

As is evident, the obtained closed form expressions are computationally
prohibitive for large scale and ill-conditioned assimilation problems
and typically first order iterative approaches (e.g., preconditioned
conjugate gradient) are required to efficiently compute the matrix
inversions. One of the main advantages of the variational formalism
is its flexibility that for example the analysis state, constrained
in a simple feasible and closed polyhedron (e.g., $\mathbf{l}\preceq\mathbf{x}\preceq\mathbf{u}$),
can be obtained using the Fast Gradient Projection (FGP) methods developed
for large scale quadratic programing problems \citep[e.g.,][]{[Nes83],[sefGL05]}.

\begin{figure}[h]
\noindent \begin{centering}
\includegraphics[scale=0.8]{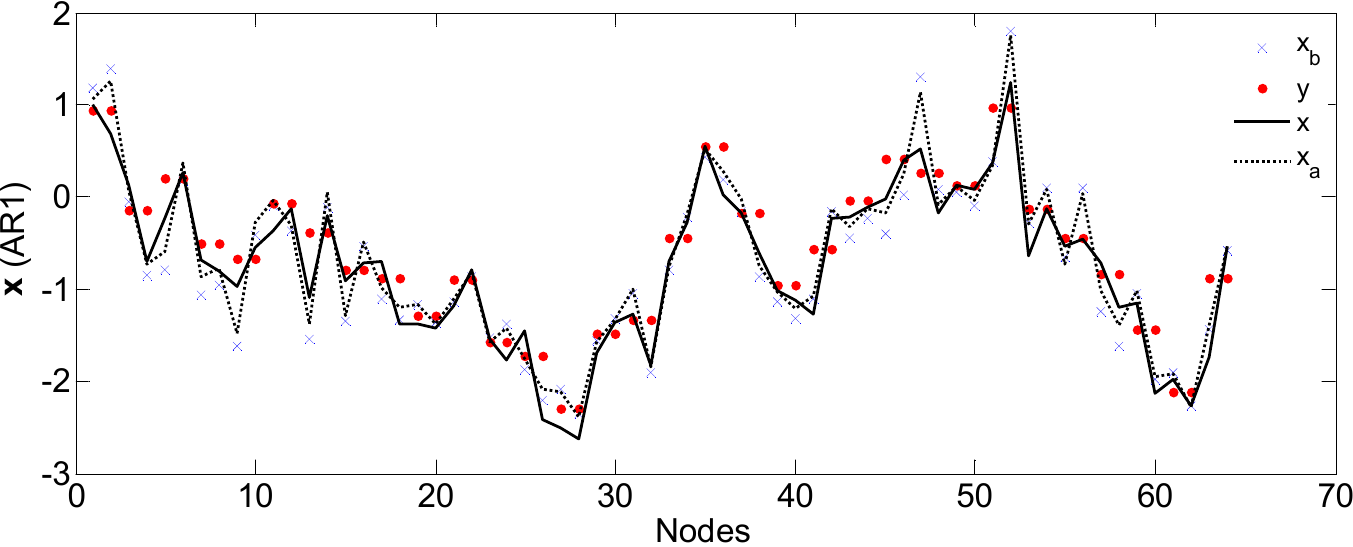}
\par\end{centering}

\caption{Results of a 3D-Var assimilation in which the true
signal ($\mathbf{x}$) follows a stationary AR(1) discrete Gaussian
process. Here we assumed $x_{i+1}=\gamma x_{i}+\sqrt{1-\gamma^{2}}u_{i}$,
where $\gamma=0.8$ and $u_{i}\sim\mathcal{N}(0,1)$. As a result
of the assimilation, the analysis exhibits an improved ${\rm SNR}=11.20$
dB.\label{fig:1} }

\end{figure}

Figure~\ref{fig:1} shows the result of a 3D-VAR assimilation scheme
applied to a stationary first order discrete Markovian process in
$\mathbb{R}^{m}$, autoregressive (AR-1), where $m=64$. A true signal
($\mathbf{x}$) is generated and the background states and observations
are obtained by adding Gaussian white noise with the signal-to-noise
ratio (SNR) $8$ dB and $10$ dB respectively, where ${\rm SNR}=20\log\left(\sigma_{\mathbf{x}}/\sigma_{noise}\right)$.
Here, to simply resemble the uncaptured subgrid details, we have assimilated
a coarse-scale observation signal with half of the size of the original
signal. To this end, we first convolved the true signal with an average
filter $1/2[+1,\,+1]$, downsampled the smoothed observations by a
factor of 2 and then added the white noise. Notice that in a matrix
form, it suffices to define the observation operator $\mathbf{H}$
as a Toeplitz convolution matrix with $h_{i,i}=h_{i+1,i}=0.5$, $h_{i,j}=0$
and then decimate the rows by a factor of 2. In other words, for each
pair of two grid points in the model space there is only one observation
node in the middle, which is assumed to be a noisy measurement of the mean
of the true states on those grid points.

In the next section, we provide evidence on the non-Gaussian and sparse
structure of the fluctuations of some important geophysical signals
in terms of their wavelet coefficients. This property is completely
ignored in the explained classical formulation of data assimilation
and can serve as additional prior knowledge to constrain more and
possibly enhance the accuracy of the assimilation results.

\section{Sparsity of Geophysical Signals}

\begin{figure}
\noindent \begin{centering}
\includegraphics[scale=0.85]{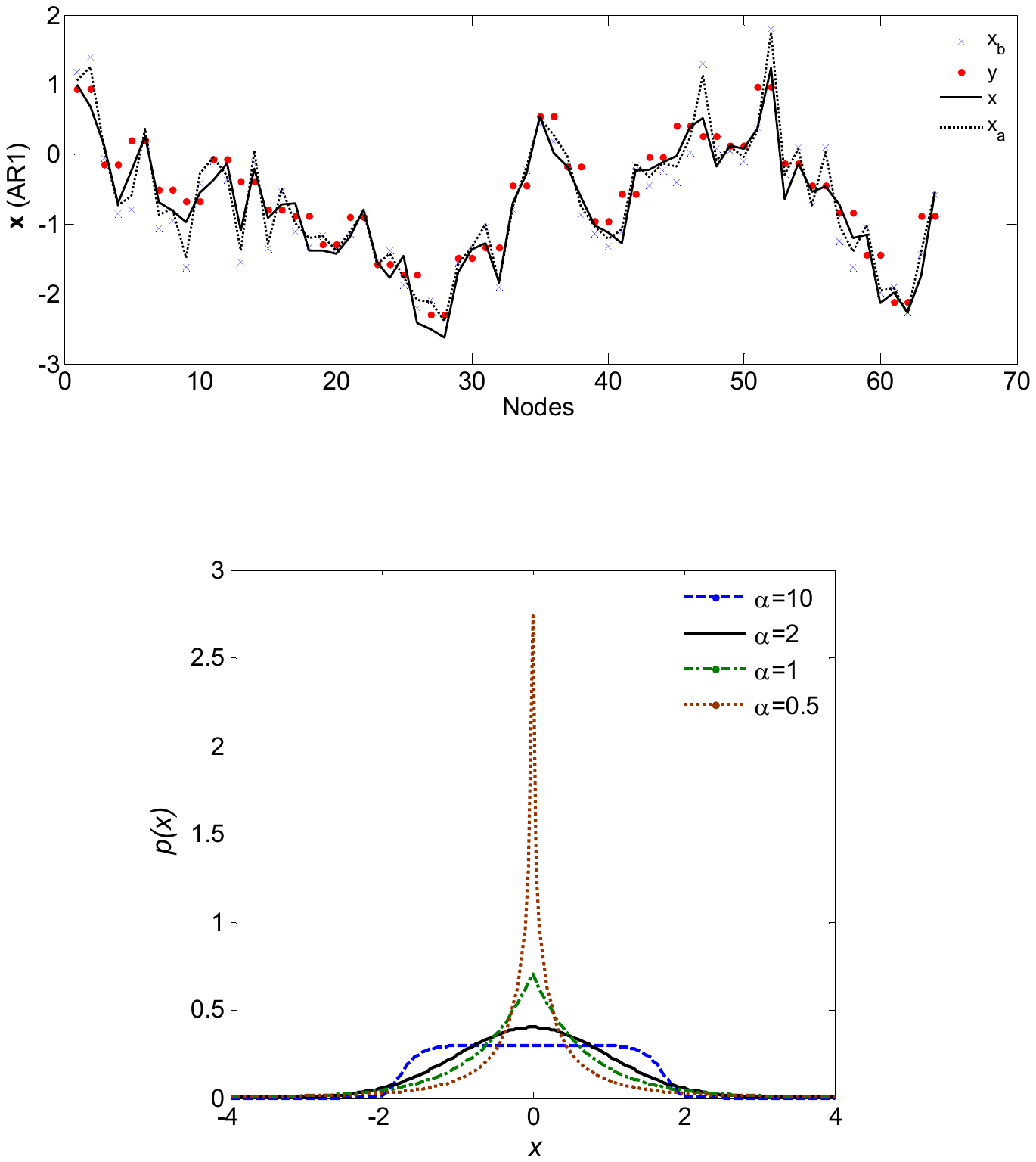}
\par\end{centering}

\caption{Generalized Gaussian Density with unit standard deviation
and various tail parameters.\label{fig:2}}
\end{figure}

Many natural signals exhibit a spatial organization of isolated high-intensity
areas nested within less active larger-scale regions. This property
often translates into a \emph{sparse representation,} that is, a major
portion of the signal can be projected onto (near) zero values under
an appropriate transformation, while only a few significantly non-zero
projection coefficients carry most of the signal energy. Motivated
by \citet{[Mal89]}, \citet{[HuaM99]} and \citet{[WaiSW01]} among
many others, it has been recently shown by \citet{[EbtF11a]} and
\citet{[EbtF12a]} that precipitation reflectivity images exhibit
a remarkably sparse representation in a redundant wavelet transform
and the distribution of the wavelet coefficients can be well explained
by the class of symmetric Generalized Gaussian Distributions (GGD).
This family of densities $p(x)\propto\exp\left(-\left|x/s\right|^{\alpha}\right)$
with a scale $s$ and tail parameter $\alpha$ spans a wide range
of exponentially bounded tail probabilities from a Dirac delta ($\alpha\rightarrow0$)
to a uniform density ($\alpha\rightarrow\infty$) in limiting cases.
The Gaussian ($\alpha=2)$ and the Laplace ($\alpha=1$) densities
are also two special cases of this family, see Figure~\ref{fig:2}.

\begin{figure*}[h]
\noindent \begin{centering}
\includegraphics[scale=1.2]{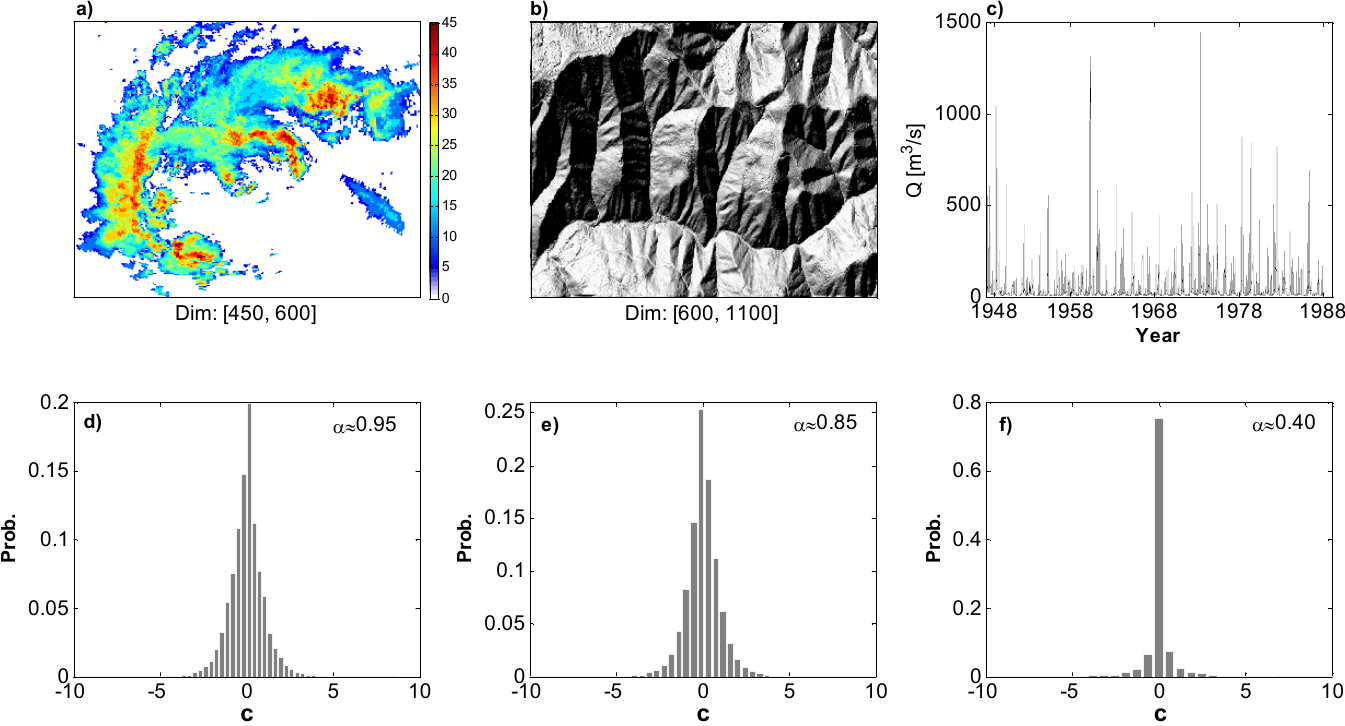}
\par\end{centering}

\caption{Sparsity of some geophysical signals, top panel from left to right:
(a) a level III NEXRAD rainfall reflectivity image in dBZ, over Texas
on 1999/03/29 (20:13:00 UTC) at resolution $\sim1\times1$ km; (b)
hillshade representation of high resolution lidar topographic data
of a small watershed (2.8 ${\rm km}^{2}$ area) in the Oregon coast
range near Coos Bay at resolution $\sim2\times2$ m; and (c) 40 years
of daily streamflow signal (1948-1988) of Leaf river basin at Collins
station (1944 ${\rm km}^{2}$ draining area), Mississippi. The bottom
panels from left to right (d)-to-(f), show the corresponding probability
histograms of the standardized wavelet coefficients $\mathbf{c}$
or say the generalized fluctuations of the above images in a probability
scale. The fitted tail parameter ($\alpha$) of the GGD is shown on
the top right corner of the lower panel plots.\label{fig:3}}
\end{figure*}

Figure~\ref{fig:3} (upper panels) shows different geophysical signals
ranging from very fast evolving dynamical processes such as precipitation
and streamflow down to a landscape digital elevation map with a very
slow evolving dynamics. Applying a Daubechies wavelet, histograms
of the wavelet coefficients share relatively similar thick tail probability
distribution, analogous to the GGD, whilst most of the values are
(near) zero; see the lower panels of Figure~\ref{fig:3}. These observations
imply that as an a priori knowledge, the wavelet coefficients (generalized
fluctuations) of these signals exhibit a sparse representation and
can be well explained, at least in part, by the family of GGDs with
the tail parameter commonly ranging in $(0,\,1]$. Note that, this
density is log-concave (i.e., the negative logarithm is a convex function)
for $\alpha\geq1$, and hence the Laplace density ($\alpha=1$) is
the best choice in this family that promotes sparsity while preserving
a convex structure. Note that, the concept of sparsity is not restricted
only to the wavelet coefficients of physical states with piece-wise
smooth structure, such as the examples presented herein. Other (prominent)
physical states with smooth surfaces and trajectories may exhibit
sparse representation in other transform domains such as the Fourier
or Discrete Cosine Transform (DCT).

\section{Assimilation with Sparse Priors}

Having informative a priori knowledge about the distribution of the
analysis state can serve to further constrain the assimilation problem
and lead to an improved a posteriori estimate of the analysis from
a Bayesian point of view. By definition, the Maximum a posteriori
(MAP) estimator of the analysis state is
\begin{equation}
\mathbf{x}_{a}^{+}=\arg\max_{\mathbf{x}}\, p\left(\mathbf{x}|\mathbf{x}_{b},\mathbf{y}\right).
\end{equation}
Applying the Bayes theorem and taking the logarithm, one can obtain
\begin{equation}
\mathbf{x}_{a}^{+}=\arg\min_{\mathbf{x}}\left\{ \mathcal{J}(\mathbf{x})-\log\, p(\mathbf{x})\right\} ,\label{eq:6}
\end{equation}
here $\mathcal{J}(\mathbf{x})$ is the negative log-likelihood function
of the classical 3D-VAR, as previously explained. Note that, for a
non-informative log-prior or say uniform density for $p(\mathbf{x})$,
this expression is exactly equivalent to the cost function of the ML estimator in equation\,(\ref{eq:3}).

A general model for the prior distribution, often referred to as the
\emph{Gibbs prior}, is given by
\begin{equation}
p(\mathbf{x})\propto\exp\left\{ -\lambda\mathcal{T}(\mathbf{x})\right\} ,\label{eq:7}
\end{equation}
where $\lambda\geq0$ is a scaling parameter and $\mathcal{T}:\,\mathbb{R}^{m}\rightarrow\mathbb{R}$
is a functional mapping from the state space to a real number, see
\citep[e.g.,][]{[ElaMR07]}. It follows from equations (\ref{eq:6})
and (\ref{eq:7}) that, the MAP estimator of the analysis state is
\begin{equation}
\mathbf{x}_{a}^{+}=\arg\min_{\mathbf{x}}\left\{ \mathcal{J}(\mathbf{x})+\lambda\mathcal{T}(\mathbf{x})\right\} ,\label{eq:8}
\end{equation}
where the non-negative $\lambda$ acts as a trade-off parameter and
plays an important role in the solution of the data assimilation problem.
Naturally, small $\lambda$ weakens the effect of sparse prior and
turns the problem into the classical least squares one, while larger
$\lambda$ values, promote a more sparse solution. In this paper,
to exploit the underlying sparsity, we suggest two particular choices
of the transformation function $\mathcal{T}(\mathbf{x})$ which yields
to a sparse-promoting reformulation of the variational data assimilation
both in the wavelet and spatial domains.

\subsection{Linear Measurement Operator}

\subsubsection{Wavelet Domain (W3D-VAR)}

Let us assume that the analysis state has a projection onto a redundant
wavelet transform $\mathbf{x}=\mathbf{\Phi}\mathbf{c}$, where the
columns of $\mathbf{\Phi}\in\mathbb{R}^{m\times k}$ contain the wavelet
``atoms'', while $\mathbf{c}\in\mathbb{R}^{k}$ is the representation
coefficients. As is evident, this matrix multiplication is equivalent
to an inverse wavelet transform while another matrix $\mathbf{\Psi}\in\mathbb{R}^{k\times m}$
represents the forward wavelet transform for obtaining the wavelet
coefficients, $\mathbf{c}=\mathbf{\Psi}\mathbf{x}$. Given that, the
wavelet coefficients of the analysis state exhibit a sparse structure
and can be well explained by independent GGD distributions, a relevant
choice for the functional mapping $\mathcal{T}(\cdot)$ in the Gibbs
prior can take the following form
\begin{equation}
\mathcal{T}(\mathbf{x})=\Sigma_{i}^{n}\left(\psi_{i}x_{i}\right)^{p}=\left\Vert \mathbf{\Psi}\mathbf{x}\right\Vert _{p}^{p}=\mathbf{\left\Vert c\right\Vert }_{p}^{p}.\label{eq:9}
\end{equation}
Choosing the closest convex representation of $\mathcal{T}(\mathbf{x})$
(i.e., $p=1$), which is equivalent to assuming a Laplace prior for
the wavelet coefficients, it follows that the 3D-VAR can be recast
in the wavelet domain as
\begin{equation}
\mathbf{c}_{a}^{+}=\arg\min_{\mathbf{c}}\left\{ \mathcal{J}(\Phi\mathbf{c})+\lambda\left\Vert \mathbf{c}\right\Vert _{1}\right\} ,\label{eq:10}
\end{equation}
where $\mathbf{c}_{a}^{+}$ denotes the analysis wavelet coefficients
that can be used to reconstruct the analysis state in the physical
state space via the inverse wavelet transform, $\mathbf{x}_{a}^{+}=\mathbf{\Phi}\mathbf{c}_{a}^{+}$.
Note that both matrix multiplications, $\mathbf{\Phi}\mathbf{c}$
and $\mathbf{\Psi}\mathbf{x}$, can be performed very efficiently
by the existing fast wavelet transforms, such as the orthogonal wavelet
transform (i.e., $\mathbf{\Phi\Psi}=\mathbf{I}$ ) \citep[e.g.,][]{[Mal89]}.
It turns out that, due to its shift invariance property, the class
of undecimated wavelet transforms is often preferred in this context,
over the traditional discrete orthogonal wavelet transform \citep[e.g.,][]{[CoiD95]}.
Notice that, assuming $p=2$ in equation (\ref{eq:9}) refers to a
Gaussian prior for the wavelet coefficients and resembles the so called
\emph{Tikhonov regularization} in solving inverse problems. In this
case, equation (\ref{eq:10}) has a closed form solution and this
choice of prior is typically very suitable for smooth states.

\subsubsection{Spatial Domain (TV3D-VAR)}

Following the existence of a sparse structure in the wavelet coefficients or say generalized fluctuations
of a geophysical signal \textbf{$\mathbf{x}$}, another choice for
the functional mapping $\mathcal{T}(\mathbf{x})$ in the Gibbs prior,
is the Total Variation (TV) semi-norm of $\mathbf{x}$, which leads
to obtaining the analysis as
\begin{equation}
\mathbf{x}_{a}^{+}=\arg\min_{\mathbf{x}}\left\{ \mathcal{J}(\mathbf{x})+\lambda\left\Vert \mathbf{x}\right\Vert _{{\rm TV}}\right\} .\label{eq:11}
\end{equation}
Two popular choices for the discrete TV semi-norm \citep[e.g.,][]{[RudOF95],[BecT09b]}
are: the isotropic one
\begin{equation}
\left\Vert \mathbf{x}\right\Vert _{{\rm TV}_{{\rm I}}}=\sum_{i=1}^{n}\sqrt{\left(\nabla_{h}x_{i}\right)^{2}+\left(\nabla_{v}x_{i}\right)^{2}},
\end{equation}
and the $l_{1}$-based
\begin{equation}
\left\Vert \mathbf{x}\right\Vert _{{\rm TV}_{{\rm l}_{1}}}=\sum_{i=1}^{n}\left(\left|\nabla_{h}x_{i}\right|+\left|\nabla_{v}x_{i}\right|\right),
\end{equation}
where, $\nabla_{h}x_{i}$ and $\nabla_{v}x_{i}$ are horizontal and
vertical first order differences at pixel \emph{i}, respectively.
Note that obtaining the optimal solution of the TV3D-VAR cost function
is more involved than the W3D-VAR as the TV semi-norm is not a separable
functional.

\subsection{Nonlinear Measurement Operator}

By first order linearization of the measurement operator in equation
(\ref{eq:2}) and a change of variable $\delta\mathbf{x}=\mathbf{x}-\mathbf{x}_{b}$,
the classical 3D-Var in an incremental form is typically reformulated
as \citep{[CouTH94]},
\begin{equation}
\mathcal{J}(\delta\mathbf{x})=\frac{1}{2}\left\Vert \delta\mathbf{x}\right\Vert _{\mathbf{B}^{-1}}^{2}+\frac{1}{2}\left\Vert \delta\mathbf{y}-\mathbf{H}\delta\mathbf{x}\right\Vert _{\mathbf{R}^{-1}}^{2},\label{eq:14}
\end{equation}
where $\delta\mathbf{y}=\mathbf{y}-\mathcal{H}(\mathbf{x}_{b})$,
and here $\mathbf{H}$ is a suitable linear approximation (e.g., Jacobian)
of $\mathcal{H}(\mathbf{x})$ in a small neighborhood around $\mathbf{x}_{b}$.
Finding $\delta\mathbf{x}_{a}$ as the minimizer of equation (\ref{eq:14}),
the analysis can be obtained by, $\mathbf{x}_{a}=\mathbf{x}_{b}+\delta\mathbf{x}_{a}$.
Having the sparse prior assumption on the transformed increments of
the analysis state, it naturally leads to a possible choice for an
incremental MAP estimator as follows:
\begin{equation}
\delta\mathbf{x}_{a}^{+}=\arg\min_{\mathbf{x}}\left\{ \mathcal{J}(\delta\mathbf{x})+\lambda\left\Vert \mathbf{\Psi}\delta\mathbf{x}\right\Vert _{1}\right\} ,\label{eq:15}
\end{equation}
where, here $\mathbf{\Psi}$ is referred to an invertible transformation
(e.g., wavelet) that sparsifies the increments. We again emphasize
the fact that the above proposed formulation seems intuitively suitable
for states with sparse increments under a proper transformation. Of
course, complementary case studies and further empirical evidence
are required for a thorough conclusion about the proper selection
of the transformation.

Notice that, although the proposed data assimilation formalism in
equations (\ref{eq:10}), (\ref{eq:11}) and (\ref{eq:15}) is convex,
the prior terms are not differentiable and hence the cost function
is \emph{non-smooth}. In this case classical (first order) gradient
based methods are no longer applicable. Several optimization techniques
have been recently proposed to deal with large-scale non-smooth convex
cost functions similar to that in equation (\ref{eq:8}), where $\mathcal{J}$
is smooth and $\mathcal{T}$ is a non-smooth convex function. A large
effort has been devoted on using efficient interior point algorithms
for this particular type of cost functions in large scale problems
\citep[e.g.,][]{[GolY05],[KimKLB07]}. However, very recently, accelerated
proximal gradient methods have received significant attention due
to their fast convergence rate and simplicity \citep[e.g.,][]{[Nes07],[FigNW07],[DiaF07],[BecT09a],[BecT09b]}.

\subsection{Non-Gaussian Error}

All of the presented formulations so far have been focused on the
fact that the model $\mathbf{v}\in\mathbb{R}^{m}$ and measurement
$\mathbf{w}\in\mathbb{R}^{n}$ error terms can be well explained by
a multivariate Gaussian distribution as the most dominant error probability
model. Sometimes the distribution of the error is symmetric with tails
markedly thicker than the Gaussian case, analogous to the GGD family.
In this case, it can be shown that naturally the ML estimator in equation
(\ref{eq:2}) is,
\begin{equation}
\underset{\mathbf{x}}{{\rm minimize}}\left\Vert \mathbf{B}^{-\nicefrac{1}{2}}\left(\mathbf{x}_{b}-\mathbf{x}\right)\right\Vert _{p_{1}}^{p_{1}}+\left\Vert \mathbf{R}^{-\nicefrac{1}{2}}\left(\mathbf{y}-\mathcal{H}(\mathbf{x})\right)\right\Vert _{p_{2}}^{p_{2}}.\label{eq:16}
\end{equation}
Notice that for $p_{i}=2$, $i\in\{1,2\}$, the problem in equation
(\ref{eq:16}) is equivalent to the classical 3D-VAR cost function
and is convex for all $p_{i}\geq1$. As explained before, it can be
concluded that the Laplace model for the error is the thickest tail
probability that can be considered while preserving convexity of the
cost function. Obviously, the above ML estimator in a Generalized
Gaussian noise environment can be further extended to the MAP estimator
by adding the prior term, as previously explained.

\section{A 1D Synthetic Example }

\begin{figure*}[t]
\noindent \begin{centering}
\includegraphics[scale=0.68]{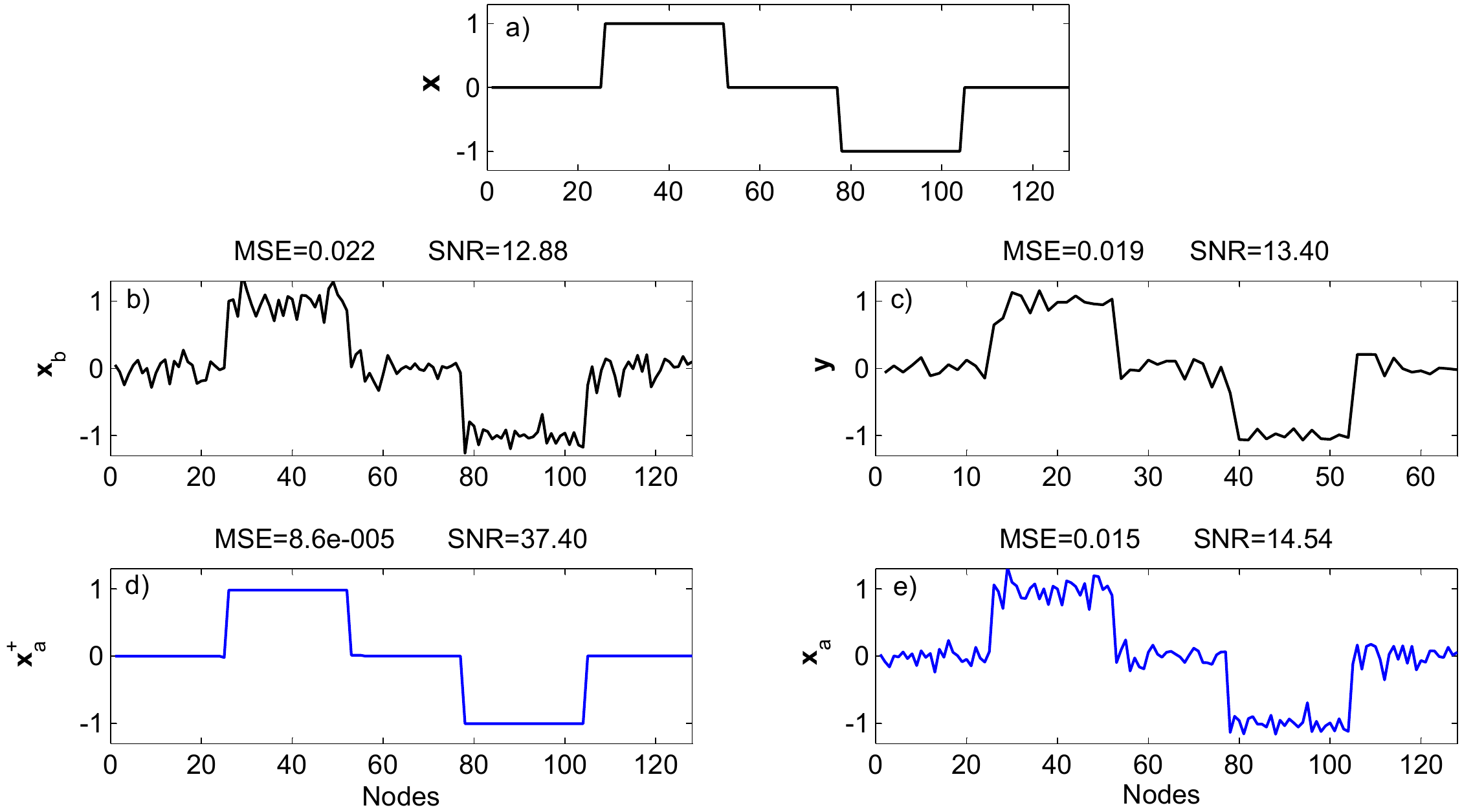}
\par\end{centering}
\caption{Demonstration of the promise of the 3D-VAR with sparse prior for a
1D example. From top-to-bottom, (a) the true signal $\mathbf{x}$,
(b) the noisy background $\mathbf{x}_{b}$, (c) the low-resolution
and noisy observation $\mathbf{y}$, (d) the obtained analysis with
prior $\mathbf{x}_{a}^{+}$ , and (e) the analysis $\mathbf{x}_{a}$
that resulted from the classical 3D-VAR. \label{fig:4}}
\end{figure*}

In this section, by no means we intend to solve a real data assimilation
problem but only to demonstrate the promise of the proposed formulations
and specifically the role of the prior on the ``analysis phase''.
To this end, we focused on a very simple piece-wise constant 1D example
with an extremely sparse structure in its first order differences,
see Figure~\ref{fig:4}. As is evident, the first order differences
of this signal exhibit a marked sparsity with only four non-zero elements
at the jump discontinuities. Notice that, in this case the wavelet
$l_{1}$-based and the explained TV-based priors become analogous
provided that the wavelet dictionaries contain the Haar ``atoms''.
To simplify and be more instructive, we have chosen a fist order differencing
operator for obtaining a sparse representation.

In this case, the 3D-VAR with sparse prior can be recast in the following
simple form
\begin{equation}
\mathbf{x}_{a}^{+}=\arg\min_{\mathbf{x}}\left\{ \mathcal{J}(\mathbf{x})+\lambda\left\Vert \mathbf{D}\mathbf{x}\right\Vert _{1}\right\} ,\label{eq:17}
\end{equation}
where, $\mathbf{D}\in\mathbb{R}^{m\times m}$ is the first order differencing
operator with $d_{i,i}=1$, $d_{i,i-1}=-1$ and $d_{i,j}=0$. To obtain
the analysis in equation (\ref{eq:17}) here we follow a quadratic
reformulation of the problem, as studied by \citet{[CheDS98]} and
\citet{[FigNW07]} and references therein. To this end, by a change
of variable $\mathbf{z}=\mathbf{D}\mathbf{x}$ one can obtain
\begin{equation}
\mathbf{z}_{a}=\arg\min_{\mathbf{x}}\left\{ \mathcal{J}(\mathbf{D}^{-1}\mathbf{z})+\lambda\left\Vert \mathbf{z}\right\Vert _{1}\right\} ,\label{eq:18}
\end{equation}
where $\mathbf{z}\in\mathbb{R}^{m}$ can be split as $\mathbf{z=u-v}$,
with $u_{i}=\max\left(z_{i},0\right)$ and $v_{i}=\max\left(-z_{i},0\right)$.
Accordingly, the $l_{1}$-prior term can be written in a linear form
as $\left\Vert \mathbf{z}\right\Vert _{1}=\mathbf{1}_{m}^{T}\mathbf{u}+\mathbf{1}_{m}^{T}\mathbf{v}$,
where $\mathbf{1}_{m}=[1,1,\ldots1]^{T}\in\mathbb{R}^{m}$. Augmenting
$\mathbf{u}$ and $\mathbf{v}$ in $\mathbf{w}=[\mathbf{u}\,\mathbf{\,\, v}]^{T}$,
equation (\ref{eq:18}) can be recast in the following constrained
quadratic programming (QP)
\begin{align}
\mathbf{\underset{\mathbf{w}}{{\rm minimize\,\,}\,\,}} & \frac{1}{2}\mathbf{w}^{T}\begin{bmatrix}\mathbf{C} & -\mathbf{C}\\
-\mathbf{C} & \mathbf{C}
\end{bmatrix}\mathbf{w}+\left(\lambda\mathbf{1}_{2m}+\begin{bmatrix}\mathbf{b}\\
\mathbf{-b}
\end{bmatrix}\right)^{T}\mathbf{w}\nonumber \\
 & \,\,\,\,\,\,\,\,\,\,{\rm subject}\,{\rm to\,\,\,\,\,\,}\mathbf{w}\succcurlyeq0,\label{eq:19}
\end{align}
where, $\lambda\geq0$, $\mathbf{C}=\mathbf{D}^{-T}\left(\mathbf{B}^{-1}+\mathbf{H}^{T}\mathbf{R}^{-1}\mathbf{H}\right)\mathbf{D}^{-1}$
and $\mathbf{b}=-\mathbf{D}^{-T}\left(\mathbf{B}^{-1}\mathbf{x}_{b}+\mathbf{H}^{T}\mathbf{R}^{-1}\mathbf{y}\right)$.
Through sub-differential analysis of equation (\ref{eq:18}), it can
be shown that for $\lambda\geq\left\Vert \mathbf{b}\right\Vert _{\infty}$
the unique minimum of equation (\ref{eq:19}) is a zero vector with
maximum possible sparsity. Here, we adopt $\lambda=0.1\left\Vert \mathbf{b}\right\Vert _{\infty}$
as suggested by \citet{[KimKLB07]}.

The example signal $\mathbf{x}\in\mathbb{R}^{128}$ ($\sigma_{\mathbf{x}}\approxeq0.65$)
is a composition of two rectangular step functions. The background
signal $\mathbf{x}_{b}\in\mathbb{R}^{128}$ in Figure~(\ref{fig:4}b)
is generated via adding a white Gaussian noise ($\sigma_{w}\approxeq0.15$,
${\rm SNR}\approxeq12.88$ dB). Here, we assimilated a low-resolution
and noisy version of the true signal as the observation $\mathbf{y}\in\mathbb{R}^{64}$
into the background signal. To this end, the observation operator
$\mathbf{H}\in\mathbb{R}^{64\times128}$ is properly designed as explained
in Section 2 and then a Gaussian white noise ($\sigma_{\mathbf{y}}=0.10$,
${\rm SNR}\approxeq13.40$ dB) is added to the downgraded version
of the true signal, see Figure~(\ref{fig:4}c). In this example,
we used the Gradient Projection with backtracking line search (i.e.,
Armijo Rule ) for solving the constrained QP in equation (\ref{eq:19}),
see \citep[p.230]{[Ber99]}. Furthermore, after obtaining $\mathbf{z}_{a}$,
we optionally recalculated the magnitude of its nonzero elements by
solely minimizing $\mathcal{J}(\mathbf{D}^{-1}\mathbf{z})$, the least
squares part of equation (\ref{eq:18}), constrained to the support
set of $\mathbf{z}_{a}$, $\mathcal{S}={\rm supp}\left(\mathbf{z}_{a}\right)$.
In other words, we assumed that the zero elements of $\mathbf{z}_{a}$
are fixed and then calculated the magnitude of its non-zero elements,
that is $\mathbf{C}_{\mathcal{S}}^{\dagger}\mathbf{b}$, where $\mathbf{C}_{\mathcal{S}}$
is a sub-matrix that contains those columns of $\mathbf{C}$ associated
to the support set $\mathcal{S}$ and $\left(\cdot\right)^{\dagger}$
is the Pseudo-inverse.

The results in Figure~\ref{fig:4} show the remarkable role of the
sparse promoting prior on the quality of the estimated analysis state.
It is clear that the estimation quality metrics have been slightly
improved by solving a classical 3D-VAR assimilation problem; however,
it led to over-fitted estimation. The result of the new proposed formulation
is close to an exact solution in this simple example and seems very
promising by outperforming the classical 3D-VAR with more than two
orders of magnitude in the SNR, which is a logarithmic metric. These
results demonstrate that the error in the observation and background
signal has been well suppressed while the discontinuities of the signal
are also well recovered due to the incorporation of the prior.

\section{Discussion and Conclusion }

We introduced a new formalism for the variational data assimilation
which takes into account a priori knowledge about the underlying statistical
structure of the state in a transformed domain and showed the preliminary
promise of the proposed methodology through a synthetic 1D example.
Although the formulation is presented for a 3D-VAR setting, it can
be extended to a 4D-VAR context. In general, we can argue that
the proper selection of the prior term typically yields a better error
(noise) suppression, while the underlying structure of the state (e.g.,
discontinuities) can also be preserved. Although the focus of this
study has been on sparsity of the state fluctuations and wavelet coefficients,
the role of the prior can be extended to other transformed domains
such as the Fourier or DCT which are intuitively more suitable for
smooth physical states.

Study of the efficient proximal gradient methods for full scale and
ill-conditioned data assimilation problems with non-smooth prior can
be of particular interest for future research in exploring the advantages
of the proposed formulations in environmental predictability. The
proposed W3D-VAR and TV3D-VAR are nonlinear estimators and hence,
estimation of the analysis error covariance is a challenge and needs
to be thoroughly investigated. As closed form expressions are not
readily available for the covariance of these nonlinear estimators,
randomization (e.g., bootstrapping) via ensemble techniques seems
a viable approach for further study in this respect.

\section*{Acknowledgments}
This work has been mainly supported by NASA-GPM award NNX07AD33G,
and an Interdisciplinary Doctoral Fellowship (IDF) of the University
of Minnesota Graduate School. The second author also wishes to acknowledge
the support provided by the Ling Chaired Professorship.


%
%

\end{document}